\documentclass[aps,pra,twocolumn,showpacs,nofootinbib]{revtex4-1}

\usepackage{amsbsy,latexsym,amsmath}
\usepackage{amsfonts}
\usepackage{amssymb}
\usepackage[mathscr]{eucal}
\usepackage{epsfig,graphics,graphicx}

\newcommand{\aver}[1]{\langle #1 \rangle}
\newcommand{\tr}{{\rm tr}}

\newcommand{\ket}[1]{\left|{#1}\right\rangle}
\newcommand{\bra}[1]{\left\langle{#1}\right|}
\newcommand{\braket}[2]{\langle{#1}|{#2}\rangle}
\newcommand{\ketbrad}[1]{\left|{#1}\rangle\!\langle{#1}\right|}

\newcommand{\be}{\begin{equation}}
\newcommand{\ee}{\end{equation}}

\newcommand{\PUA}{P^{\mathrm ?}}
\newcommand{\PME}{P^{\mathrm{ME}}}

\newcommand{\na}{n_A}
\newcommand{\nb}{n_B}
\newcommand{\nc}{n_C}

\begin{document}
\title{Multi-copy programmable discrimination of general qubit states}
\author{G. Sent\'{\i}s, E.~Bagan, J.~Calsamiglia and R.~Mu\~{n}oz-Tapia}
\affiliation{F\'{\i}sica Te\`{o}rica: Informaci\'{o} i Fen\`{o}mens Qu\`{a}ntics, Facultat de
Ci\`{e}ncies, Edifici Cn, Universitat Aut\`{o}noma de Barcelona, 08193
Bellaterra (Barcelona) Spain}

\begin{abstract}

Quantum state discrimination is a fundamental primitive in quantum statistics where one has to correctly identify
the state of a system that is in one of two possible known states. A programmable discrimination machine performs this task when the pair of possible states is not a priori known, but instead the two possible states are provided through two respective program ports.
We study optimal programmable discrimination machines for general qubit states when several copies of states are available in the data or program ports. Two scenarios are considered: one in which the purity of the possible states is a priori known, and the fully universal one where the machine operates over generic mixed states of unknown purity. We find analytical results for both, the unambiguous and minimum error, discrimination strategies. This allows us to calculate the asymptotic performance of  programmable discrimination machines when a large number of copies is provided, and to recover the standard state discrimination and state comparison values as different limiting cases.

\end{abstract}
\maketitle
\section{Introduction}

Discrimination between given hypotheses is one the most basic tasks in our every day lives. Very often we are confronted with the necessity of having to  identify an option between some possible choices based on some acquired evidence.
In  the quantum setting the discrimination problem consists of identifying one of two possible states given a number of identical copies available for measurement.
This task  encompasses a plethora of non-trivial theoretical and experimental implications. In the usual setting the a priori states are known, i.e., the classical information characterizing the possible states is provided and the discrimination protocol is tailored for this specific information. One usually considers two types of approaches:
unambiguous~\cite{unambiguous} and minimum error~\cite{helstrom} discrimination. An unambiguous protocol is one where the identification of the state
is error free. Of course,  this is only possible stochastically, i.e, unless the states are orthogonal, the protocol must give an inconclusive answer (the ``I do not know'' outcome) with a non vanishing probability. In the minimum error approach, the protocol always yields a definite answer, which may be wrong some of the times.
An optimal protocol is one which minimizes the inconclusive or the error probability. It may also be possible to go continuously from one case to the other by considering margins of error probabilities~\cite{continuous}.  In spite of being such a fundamental problem, only very recently a closed expression for the asymptotic error probability has been obtained (see~\cite{chernoff, nussbaum, chernoff-locc} and references therein), the quantum Chernoff bound, from which metric distances and state densities~\cite{geometry} can be derived.

Very much in the spirit of universal computers, it is interesting to consider discrimination devices that are not specialized in a specific discrimination instance but can discriminate between arbitrary pairs of states \cite{dusek,bergou-hillery}.  In these,  the set of possible states enter the device as ``programs", i.e., the classical description of the states is not provided beforehand, rather the information is incorporated in a quantum way (this can also be viewed as an instance of relative information~\cite{relative}).  These devices have program ports that are loaded with the program states,  and a data port that is loaded with the unknown input state one wishes to identify.  The device will identify the state of the data port as being one of the  states fed in the program ports, but this identification will in general be erroneous with a probability that decreases with the number of copies of the states entering the ports.  One can also regard these devices as learning machines~\cite{learning}, where the device is instructed through the program ports about  different states, and based on this knowledge the machine
associates the state in the data port with one of the states belonging to the training set.   Increasing the number of copies of states at the program and data ports of course increases the chances of correct identification.

It is particularly relevant to understand how the probability of error scales with an increasing number of copies and what are the corresponding error rates. The value of this rate is one of the most relevant parameters assessing the performance of the device. We will consider  the discrimination of two general qubit states, although most of our results can be generalized to higher dimensional systems (see \cite{coherent,coherent-exp} for a single copy continuous variable setting). For simplicity we will assume that the prior occurrence probability of each state is identical and compute the unambiguous and minimum error rates for optimal programmable devices.

We first study the performance of such devices for pure states.  We compute the  error probabilities  for any number of pure qubit states at the input ports. Some of the results are already available in the literature~\cite{bergou-hillery,bergou-feldman-hillery,hayashi,he-bergou,bergou-buzek,he-bergou,qudits},
but the way we formalize the problem here is crucial to treat the more general mixed state case. In addition we obtain analytical expressions that enable us to present the results and study limiting cases in a unified way. In particular, when the program ports are loaded with an infinitely large number of copies of the states  we recover the usual state discrimination problem~\cite{helstrom}, since it is clear that then one has the classical information determining  the states entering the program ports. On the other hand, when the number of copies at the data port is infinitely large, while the number  of copies at the program ports are kept finite, we recover the state comparison problem~\cite{comparison,sedlak}.

We extend the previous pure state study to the case of mixed input states. In this scenario we only compute the minimum error probability, as no unambiguous answers can be given if the states have the same support~\cite{mixed-ua}. The performance of the device for a given purity of the input states allows to quantify how the discrimination power is degraded in the presence of noise. The expressions here are much more involved, however one can still exploit the permutation symmetry of the input states to write the problem in a block-diagonal form~\cite{deconstruction,deconstruction-2}. We then obtain closed expressions for the  probability of error that can be computed analytically for small number of copies and numerically evaluated for a fairly large number of copies. We are also able to  obtain analytical expressions for some  asymptotic rates. Again, the leading term, as in the pure state case, is seen to coincide with the average minimum error for known states.


We also analyze the fully universal discrimination machine, i.e.,  a device that works optimally for completely unknown input states. In this case one has to
assume a uniform distribution for the purity. In contrast to the pure state distribution,  there is no unique choice~\cite{petz}, and different reasonable assumptions
lead to different uniform priors. Here we consider hard-sphere, Bures and Chernoff priors.  

The paper is organized as follows. In the next Section we obtain the error probabilities for pure states  when each program port is fed with $n$ copies of each state and there are $m$ copies of the unknown state entering the data port. In Sec~\ref{sec:pure-limits} we study the asymptotic rates in several scenarios.
In Section~\ref{sec:mixed} we analyze the performance of these devices when the ports are loaded with copies of  states of known purity and  obtain some interesting limiting cases in Sec.~\ref{sec:mixed-limits}.  We finally obtain the error rates for the  fully universal programmable machine. Some brief conclusions follow and we end up with two technical appendices.

\section{Pure states}\label{sec:pure}
Let us start by fixing the notation and conventions used throughout this paper. We label the two program ports by $A$ and $C$. 
They will be loaded with states $\ket{\psi_1}$ and $\ket{\psi_2}$, respectively. The data port, $B$, is the middle one
and will be loaded with the states we wish to identify. We also use the short hand notation $[\psi]$ to denote $\ketbrad{\psi}$  and similarly
$[\psi \phi\ldots] =[\psi]\otimes[\phi]\otimes\cdots=\ketbrad{\psi}\otimes \ketbrad{\phi}\otimes\cdots$.  We may also omit the subscripts  $A, B$ and $C$ when no confusion arises. We assume that the program ports are fed with $n$ copies of each state and the data port with $m$ copies of the unknown state.
This is a rather general case for which closed expressions of the error probabilities can be given. The case with arbitrary $n_A, n_B,$ and $n_C$ copies at each port is discussed in appendix~\ref{na-nb-nc}. The expressions are more involved but the techniques are a straightforward extension of the ones presented here.

When  the state at the data port is $\ket{\psi_1}^{\otimes m}$ or  $\ket{\psi_2}^{\otimes m}$, the effective states entering the machine are given by the averages
\begin{eqnarray}\label{multi-copy-pure}
\sigma_1&=&\int d\psi_1 d\psi_2 [\psi_1^{\otimes n}]_A [\psi_1^{\otimes m}]_B [\psi_2^{\otimes n}]_C \nonumber \\
\sigma_2&=&\int d\psi_1 d\psi_2 [\psi_1^{\otimes n}]_A [\psi_2^{\otimes m}]_B [\psi_2^{\otimes n}]_C\, ,
\end{eqnarray}
respectively. %
The integrals can be easily computed using Schur lemma, $\int d\phi [\phi]_X = \frac{1}{d_X} \openone_X$, where $d_X$ is the dimension of the
Hilbert space spanned by $\{\ket{\phi}\}$ and $\openone_X$ is the projector onto this space. Hence
\begin{eqnarray}\label{rho1-rho2}
		\sigma_1 &=&    \frac{1}{d_{AB} d_C} \openone_{AB} \otimes \openone_{C} \nonumber \\
		\sigma_2 &=&    \frac{1}{d_{A} d_{BC}} \openone_{A} \otimes \openone_{BC}\, ,
\end{eqnarray}
where $ \openone_{XY} $ is the projector onto the completely symmetric subspace of $\mathcal{H}_X\otimes \mathcal{H}_Y$
and $d_{XY}=\tr\openone_{XY}$ is its dimension. For qubits we have $d_A=d_C=n+1$ and $d_{AB}=d_{BC}=n+m+1$.

The structure of the states \eqref{rho1-rho2} suggests the use of the angular momentum basis, $\ket{j_A,j_B (j_{AB}), j_{C};J M}$ for
$\sigma_1$ and $\ket{j_{A},j_B,j_C (j_{BC});J M}$ for $\sigma_2$. The quantum numbers $j_{AB}=j_A+j_B$ and $j_{BC}=j_B+j_C$  recall the way
the three spins are coupled to give the total
angular momentum $J$. Here the angular momenta have a fixed value determined by the number
of copies at the ports, $j_A=j_C=n/2$, $j_B=m/2$.  So, we can very much ease the notation  by only
writing explicitly the  labels $j_{AB}$ and $j_{BC}$.
We would like to stress however that in general one needs to  keep track of all the quantum numbers, specially when dealing with mixed states as in
Sec.~\ref{sec:mixed}.

In $\sigma_1$ the first $n+m$ spins are coupled in a symmetric way, while in  $\sigma_2$  the symmetrized spins are the last $n+m$, thus
$j_{AB}=(n+m)/2=j_{BC}$.  The states are diagonal in the angular momentum bases discussed above, and we have
\begin{eqnarray}
\sigma_1&=& \frac{1}{d_{AB} d_C}\sum_{J=0,1/2}^{m/2+n} \sum_{M=-J}^{J} [j_{AB};J M]\nonumber \\
\sigma_2&=& \frac{1}{d_A d_{BC}}\sum_{J=0,1/2}^{m/2+n} \sum_{M=-J}^{J} [j_{BC};J M]\, ,
\end{eqnarray}
where the lower limit of the first summation takes the value 0 (1/2) for $m$ even (odd).
Notice that the spectrum of both matrices is identical and that the basis elements of their support
differ only in the way the three spins are coupled. Further, the key feature of the total angular momentum bases is the orthogonality relation
\begin{equation}\label{orthogonality}
\braket{j_{AB} ;J M}{j_{BC} ;J' M'} =0  \quad \forall J \neq J' \mathrm{or}  M\neq M' .
\end{equation}
Bases of this type are known as Jordan bases of subspaces~\cite{bergou-feldman-hillery}. Since a state of the first basis (labeled by $ j_{AB} $) has overlap with only one state of the second basis (labeled by $ j_{BC} $),
the problem is reduced to a discrimination instance between pairs of pure states.
Then the total error probability is simply the sum of the contributions of each pair.

In the unambiguous approach, the minimum probability of an inconclusive result
for a pair of states $\ket{\phi_1},\ket{\phi_2}$  with equal priors  is simply  $\PUA(\ket{\phi_1},\ket{\phi_2})=|\braket{\phi_1}{\phi_2}|$ \cite{unambiguous}, hence
\begin{equation}\label{inconclusive}
\PUA=\frac{1}{d_{AB} d_{C}}\sum_{JM}|\braket{j_{AB} ;J M}{j_{BC} ;J M}|.
\end{equation}
These overlaps can be computed in terms of the Wigner 6-$j$ symbols~\cite{edmonds}:
\begin{widetext}
\begin{equation}\label{6j}
\braket{j_{AB}; JM}{j_{BC}; JM}= (-1)^{j_A+j_B+j_C+J} \sqrt{(2j_{AB}+1)(2j_{BC}+1)}
                                                   \begin{Bmatrix} j_A & j_B & j_{AB} \\ j_C & J & j_{BC} \end{Bmatrix}\, ,
\end{equation}
\end{widetext}
and they are independent of  $M$~\cite{edmonds}, therefore in what follows we omit writing the quantum number $M$, and we perform the sum over $M$ in \eqref{inconclusive}  trivially by adding the multiplicative factor
$2J+1$.  Substituting the value of the $6j$ symbols for $j_A=j_C=n/2$, $j_B=m/2$, $j_{AB}=j_{BC}=(n+m)/2$, and setting $J=m/2+k$ we obtain
\begin{equation}\label{overlap-nm}
\braket{j_{AB}; J}{j_{BC}; J}=
 \displaystyle \binom{n}{k} \binom{n+m}{n-k}^{-1}\, ,
\end{equation}
with $k=0,1,\ldots,n$ (observe that $J$ takes values from $J=n+m/2$ of the totally symmetric space down to $J=m/2$).

Plugging the overlaps Eq.~\eqref{overlap-nm}  into Eq.~\eqref{inconclusive}, we obtain
\begin{eqnarray}
\PUA&=&\sum_{k=0}^{n} \frac{m+2k+1}{(n+m+1)(n+1)}\frac{(m+k)!n!}{(m+n)!k!}\nonumber \\
 &=&1- \frac{n m}{(n+1)(m+2)}, \label{ua-nm}
\end{eqnarray}
where notice that the dimension of the subspace of total angular momentum $J$ is $m+2k+1$ and in the second equality we have used the binomial sums
\begin{eqnarray}
\sum_{k=0}^n {m+k \choose m}&=&{n+m+1 \choose m+1}\, ,\nonumber \\
\sum_{k=0}^n k {m+k \choose m}&=&{n+m+1 \choose m+1} \frac{n(m+1)}{m+2} .
\end{eqnarray}

In the minimum error approach no inconclusive results are allowed, but the  machine  is permitted to give wrong answers with some probability
that one tries to minimize. This minimum error probability can be computed along the same lines as in the previous case.
Recall that the error probability $\PME$ for two pure states $\ket{\phi_1},\ket{\phi_2}$  and equal a priori probabilities is \cite{helstrom}
\begin{equation} \label{minerr}
		\PME(\ket{\phi_1},\ket{\phi_2})=  \frac{1}{2} \left( 1 - \sqrt{1-|\braket{\phi_1}{\phi_2}|^2} \right) \; .
\end{equation}
The total error probability is just the sum of the contribution of each pair of states  with the same quantum numbers $JM$, $\{\ket{j_{AB}; JM}, \ket{j_{BC}; JM}\}$,
\begin{eqnarray}\label{min-nm}
		\PME&=&\frac{1}{2} \left( 1-\sum_{k=0}^n \frac{m+2k+1}{(n+1)(n+m+1)} \right. \nonumber \\
		&&\phantom{xxxxxxxx} \times \left.\sqrt{1-\left(\frac{(m+k)!n!}{(m+n)!k!}\right)^2} \right) \, .
		\end{eqnarray}
%

It is instructive to obtain the well-known results when the ports are loaded with just one copy of each state~\cite{bergou-hillery}, i.e., $n=m=1$.

The inconclusive probability in the unambiguous approach reads
\begin{eqnarray}
\PUA &=&\frac{1}{6}\sum_{J=1/2}^{3/2} (2J+1) |\braket{j_{AB}=1 ;J }{j_{BC}=1; J }|\nonumber \\
        &=&\frac{5}{6},
\end{eqnarray}
i.e.,  five out of six times the machine gives an inconclusive result and only 1/6 of the times identifies the state without error. Notice that the
overlaps for $J=3/2$ are one. This must be so since $J=3/2$ corresponds to
the totally symmetric subspace, which is independent of the way the spins coupled. That is, this subspace is identical for $\sigma_1$ and $\sigma_2$. This is the main source of error as it contributes $4/6=4/6\times 1$  out of the total $5/6$ error probability. The remaining $1/6=2/6\times 1/2$ is the  contribution of the
$J=1/2$ subspace, where the $2/6$ is the probability of having an outcome on this subspace and $1/2$ is the overlap between the states [cf. Eq.~\eqref{overlap-nm}].

The minimum error probability in the one copy case reads
\begin{eqnarray} \label{minerr-tot}
		\PME&=&  \frac{1}{2} \left( 1 - \frac{1}{6} \sum_{J=1/2}^{3/2} (2J+1)\right.\nonumber \\
		&& \phantom{xx}\times   \sqrt{1- |\braket{j_{AB}=1 ;J }{j_{BC}=1; J }|^2} \Bigg) ,
\end{eqnarray}
which using Eq.~\eqref{overlap-nm} or directly Eq.~\eqref{min-nm} gives
\begin{equation} \label{minerr-tot-2}
		\PME =  \frac{1}{2} \left( 1 - \frac{1}{2\sqrt{3}} \right)\simeq 0.356 \; .
\end{equation}
That is, approximately 1/3 of the times the outcome of the machine will be incorrect.

The error probability in both, minimum error and unambiguous approaches, will of course decrease when using more copies of the states at the ports
of the discrimination machine.
  Equations \eqref{ua-nm} and \eqref{min-nm} give the unambiguous and minimum error
probability for arbitrary values of $n$ and $m$. They enable us to study  the behaviour of the machine for a large
number of copies in the program and the data ports, which is what we next discuss.

\section{Asymptotic limits for pure states}\label{sec:pure-limits}
Let us start by considering the case of an asymptotically large number of copies at the program ports ($n \to \infty $) while keeping finite the number of copies $m$ at the data port.
For the unambiguous discrimination one obtains from Eq.~\eqref{ua-nm}
\begin{equation}\label{SUA-limit-n}
	\lim_{n\to\infty} \PUA=\frac{ 2}{m+2}\, .
\end{equation}
We wish to show that in this limit the programmable machine has a performance that
is equivalent to a protocol consisting in first estimating the states and then doing a discrimination of \emph{known} states. The average of the  inconclusive  probability of this protocol over all input states should coincide with Eq.~\eqref{SUA-limit-n}.  Recall that for known $\ket{\psi_1}$ and $\ket{\psi_2}$ states, when
a number $m$ of copies of the unknown state is given, this probability reads
\begin{equation}\label{SUA-known}
	 \PUA (\psi_1,\psi_2)=\left|\braket{\psi_1}{\psi_2}\right |^m \, .
\end{equation}
One can do an explicit calculation of the average $\aver{ \PUA (\psi_1,\psi_2)}=1/2\int_0^{\pi} \sin\theta \cos^m\theta/2$, but it is amusing to obtain it
in a very simple way from the
Schur lemma
\begin{eqnarray}\label{SUA-known-average}
	 \int d\psi_2 \left(\left|\braket{\psi_1}{\psi_2}\right |^2\right)^{\frac{m}{2}} &=&\bra{\psi_1}  ^{\otimes \frac{m}{2}}\left(\int d\psi_2 [\psi_2]^{\otimes \frac{m}{2}}\right) \ket{\psi_1}^{\otimes\frac{m}{2}} \nonumber \\
	 &=& \frac{1}{d_{m/2}}=\frac{1}{m/2+1}\, ,
\end{eqnarray}
where $d_{m/2}$ is the dimension of the symmetric space of $m/2$ qubits (notice that stricto sensu this procedure is only
valid for $m$ even). Plugging this average into \eqref{SUA-known} one immediately recovers Eq.~\eqref{SUA-limit-n}.

Now we turn our attention to the minimum error probability.  Taking $n \rightarrow \infty$  and using the Stirling approximation
$z!\approx z^z \mathrm{e}^{-z}\sqrt{2 \pi z}$ in Eq.~\eqref{min-nm}, one obtains
\begin{eqnarray}\label{SME-sum}
\lim_{n\to\infty} \PME&=&\frac{ 1}{2}\left[ 1-2\int_0^1 dx \, x \sqrt{1-x^{2m}}  \right] \nonumber \\
&=&\frac{1}{2}\left[  1-\frac{\sqrt{\pi}}{2}\frac{\Gamma(1+1/m)}{\Gamma(3/2+1/m)} \right]\, ,
\end{eqnarray}
where we have defined $x=k/n$ and used the Euler-McLaurin summation formula at leading order $\sum_{k=0}^n f(k)\simeq n \int_0^1 dx f(n x)$.

This result could be easily anticipated  from the minimum error probability with classical knowledge of the  pure states.
Recall that  the minimum error probability
given  $m$ identical copies is
\begin{equation}
\PME(\psi_1,\psi_2)=\frac{1-\sqrt{1-|\braket{\psi_1}{\psi_2}|^{2m}}}{2},
\end{equation}
and we just have to compute the average for all pairs of the above expression.  Using $|\braket{\psi_1}{\psi_2}|^2= \cos^2\theta/2$, where $\theta$ is the relative angle between the Bloch vectors of the two states, one has
\begin{equation}\label{PME-int}
\aver{\PME(\psi_1,\psi_2)}=\frac{1}{2}\left[ 1-\frac{1}{2} \int_0^\pi d\theta \sin\theta \sqrt{1-\cos^{2m}(\theta/2)}\right],
\end{equation}
and performing the change of variables $x=\sin\theta/2$  this equation is cast exactly in the form of \eqref{SME-sum}.

What can not be anticipated is the next order $O(1/n)$, which  gives the very relevant information on how fast the protocol
reaches the asymptotic value~\eqref{SME-sum}.  A lengthy, but rather straightforward, calculation yields the remarkable result that this term has a coefficient
which coincides with the value of the  integral $\int_0^1 dx \, x \sqrt{1-x^{2m}}$. At this order we therefore can write
\begin{equation}\label{SME-subleading}
\PME=\frac{1}{2}-\frac{\sqrt{\pi}}{4}\frac{\Gamma(1+1/m)}{\Gamma(3/2+1/m)} \left(1-\frac{1}{n}\right).
\end{equation}

We now analyze the complementary case, that is,  when the number of copies at the data port is infinitely large, $m\to \infty$, while the number $n$  of copies at the program ports is kept finite. In this limit we have perfect knowledge of the data state, $\ket{\psi}$, but we do not know to which program port it should be associated. Observe that this situation is very much the same as state comparison~\cite{comparison}.

In this scenario the inconclusive probability in the unambiguous approach reads  from~\eqref{ua-nm}
\begin{equation}\label{SUA-limit-m}
	\lim_{m\to\infty} \PUA=\frac{1}{n+1} \, .
\end{equation}
Let us see that this agrees with the average performance of standard state comparison.  If the data state is the same as the program state in the  upper or lower port,
the effective states to be discriminated  are
\begin{eqnarray}\label{state-minfinity}
  \sigma_1&=&\frac{1}{d_n}[\psi^{\otimes n}]\otimes\openone_n \nonumber \\
  \sigma_2&=&\frac{1}{d_n}\openone_n\otimes[\psi^{\otimes n}] \, ,
\end{eqnarray}
 respectively,
  where $d_n=n+1$ is the dimension of the symmetric space of $n$-qubits and $\openone_n$ is the projector onto this subspace. The minimal inconclusive probability for these two states can be obtained with a POVM whose elements are $\{E_1= [\psi^{\otimes n}]\otimes[\psi^{\otimes n}]^{\bot} ,E_2=[\psi^{\otimes n}]^{\bot}\otimes[\psi^{\otimes n}] ,E_?=\openone \otimes \openone -E_1 -E_2  \}$,
where $[\psi^{\otimes n}]^{\bot}=\openone_n-[\psi^{\otimes n}]$,  that is,  with a  POVM that checks in each register wether the state is  $\ket{\psi}$ or not . Then the inconclusive probability reads
\begin{equation}\label{eq:UA-averaged}
\PUA(\psi)=\frac{1}{2}\left(\tr E_? \sigma_1+ \tr E_? \sigma_2 \right)=\frac{1}{n+1}
\end{equation}
independently of the state $ \ket{\psi} $, where we have used $\tr [\psi^{\otimes n}]^{\bot}=\tr (\openone_n-[\psi^{\otimes n}])= n$.

The minimum error probability in this limit can be tackled in a similar fashion. The asymptotic expression of
Eq.~\eqref{min-nm}, though not as direct as in the unambiguous case,  is rather straightforward to obtain.
Notice that the dominant factor in the  term containing factorials inside the square root is $m^{-2(n-k)}$. So, we can effectively replace the
square root term  by one, for all $k<n$. Taking into account that for  $k=n$ the square root vanishes, we have
\begin{equation}\label{SME-limit-m}
	\lim_{m\to\infty} \PME=\frac{ 1}{2}\left( 1- \frac{n}{n+1}\right)=\frac{1}{2(n+1)} \, .
\end{equation}

The minimum error probability of  a strategy that first estimates perfectly the input state and then tries to associate the correct label to it is given by
       Helstrom formula for $\sigma_1$ and $\sigma_2$  \cite{helstrom}
\begin{eqnarray}\label{Helstrom-m}
\PME  &=&\frac{1}{2}(  1-\frac{1}{2} \|\sigma_1-\sigma_2 \|)\, ,
\end{eqnarray}
where $\| A\|=\tr \sqrt{A^\dag A}$ is the trace-norm of operator $A$. Substituting the expression of the states \eqref{state-minfinity} we obtain
\begin{eqnarray}
    \PME     &=&\frac{1}{2}\left( 1-\frac{1}{2(n+1)} \| [\psi^{\otimes n}]
         \otimes [\psi^{\otimes n}]^{\bot} \right. \nonumber \\
         &&\phantom{xxxxxxxxxxx} -   [\psi^{\otimes n}]^{\bot}\otimes[\psi^{\otimes n}] \| \bigg)\nonumber \\
          &=&\frac{1}{2}\left(  1-\frac{2}{2(n+1)} \| [\psi^{\otimes n}]\otimes [\psi^{\otimes n}]^{\bot}\| \right)\nonumber \\
          &=&\frac{1}{2}\left(  1-\frac{n}{n+1} \right)=\frac{1}{2(n+1)},
          \label{eq:ME-averaged}
\end{eqnarray}
where in the first equality we have subtracted the common term $[\psi^{\otimes n}]\otimes[\psi^{\otimes n}]$ from both states,
in the second we have used the orthogonality of the operators and in the last equality  we use  $\tr [\psi^{\otimes n}]^{\bot}=n$ . As expected, the result is again independent of $\ket{\psi}$.

To end this section we compute the asymptotic error probabilities for the symmetric case, that is, when all the ports are loaded with the same $m=n$ (and large) number of copies.

In the unambiguous approach when $n=m\to \infty$ the first nonvanishing order of~\eqref{ua-nm} reads
\begin{equation}
\PUA=\frac{3}{n}+\ldots
\end{equation}

To compute the minimum error probability, it is convenient to write Eq.~\eqref{min-nm} for $n=m$ as
\begin{equation}
\PME=\frac{1}{2}\sum_{k=0}^n p_k \left(1- \sqrt{1-c_k^2}\right)\, ,
\end{equation}
where
\begin{equation}\label{pk}
p_k=\frac{n+1+2k}{(2n+1)(n+1)}
\end{equation}
and
\begin{equation}
c_k=\frac{\binom{n+k}{n}}{\binom{2n}{n}}\, .
\end{equation}
We first observe that $c_k$ is a monotonically increasing function and hence it takes its maximum value  at $k=n$. Second,
we note that around this point
\begin{eqnarray}
\binom{n+k}{n}&\simeq& 2^{(n+k) H(\frac{n}{n+k})}\nonumber \\
&\simeq& 2^{(n+k) H(1/2)}=2^{n+k}\, ,
\end{eqnarray}
where $H(x)=-x\ln x-(1-x)\ln(1-x)$ is the Shannon entropy of a binary random variable and we have used that $k\approx n$ and $H(1/2)=1$.
Similarly, one has
\begin{equation}
\binom{2n}{n} \simeq 2^{2n H(1/2)}=2^{2n}\,
\end{equation}
and hence $ c_k\simeq 2^{-(n-k)} $. With this, the probability of error in this limit reads
\begin{equation}
\PME=\frac{1}{2}\sum_{k=0}^\infty p_k \left(1- \sqrt{1-\left(\frac{1}{4}\right)^{n-k}}\right)\,.
\end{equation}
Finally, we perform the change of variables $k\to n-k$ and use that in Eq.~\eqref{pk} $p_{n-k}\simeq 3/(2 n)$ for $k\simeq 0$ to obtain
\begin{equation}\label{pure-assymptotic}
\PME=\frac{3}{4n}\zeta(1/4)\approx \frac{0.882}{n}\, ,
\end{equation}
where we have defined the function
\begin{equation}
\zeta(x)=\sum_{k=0}^\infty\left(1-\sqrt{1-x^k}\right)\, ,
\end{equation}
which converges very quickly to its exact value (the first  four terms already give a value that differ in less than $10^{-3}$ from the exact value).


\section{Mixed states}\label{sec:mixed}
We now move to the case when the program and data ports are loaded with mixed states. This situation arises when, e.g., there are imperfections in the preparation or noise in the transmission of the states.  It is reasonable to suppose that these  imperfections have the same effect on all states, i.e. to consider that the states have all the same purity $r$. The input states are then tensor products of
\begin{equation}\label{mixed-qubit}
\rho_i=\frac{\openone + r \vec{n}_i \vec{\sigma}}{2} \ ,
\end{equation}
where $\vec{n}_i$ is a unitary vector and $\vec{\sigma}=(\sigma_x,\sigma_y,\sigma_z)$ are the usual Pauli matrices.  In what follows we assume that only
the purity is known, i.e. one knows the characteristics of  the noise affecting the states, but nothing else. This  means that the averages will be performed over the
isotropic Haar measure of the $\mathbb{S}^2$ sphere, like for pure states. At the end of this section we also analyze the performance of a fully universal discrimination machine, that is, when not even the purity is considered to be known.

Notice  that mixed states can only be unambiguously discriminated if they have different supports \cite{mixed-ua}, which is not the case when the ports are loaded with copies of the states \eqref{mixed-qubit}  as they are  full-rank matrices.
Therefore, only the minimum error discrimination approach will be analyzed here. It is worth stressing that the computation of the optimal error probability in the multi-copy case is very non-trivial, even for known qubit mixed states. Only recently feasible methods for computing the minimum error probability for a rather large number of copies have been developed and the  asymptotic expression of this probability has been obtained~\cite{chernoff,chernoff-locc}. The main difficulty  can be traced back to the computation of the trace-norm [see Eq.\eqref{Helstrom-m}] of large matrices. The dimension of the matrices grows exponentially with the total number of copies entering the machine, and for a relative small number of them the problem becomes unmanageable. However, as it will be clear, it is possible to exploit the permutation symmetry of the input states to write them in block-diagonal form ~\cite{deconstruction,deconstruction-2} crucially reducing the complexity of the problem.

The two effective states we have to discriminate are
\begin{eqnarray}\label{s1-s2-int}
\sigma_1 &= & \int dn_1 dn_2 \rho^{\otimes n}_{1\,  A} \otimes \rho_{1\,  B}^{\otimes m} \otimes \rho^{\otimes n}_{2\,  C}  \nonumber \\
\sigma_2 &=&  \int dn_1 dn_2 \rho^{\otimes n}_{1\,  A} \otimes \rho_{2\,  B}^{\otimes m} \otimes \rho^{\otimes n}_{2\,  C} ,
\end{eqnarray}
where $dn_i= d\Omega_i/(4\pi)$ is the invariant measure on the two-sphere. Any state having permutation invariance, as e.g.  $ \rho^{\otimes n} $,  can be written in a
block diagonal form using the irreducible representations of the symmetric group $S_n$. Each block is specified by the total angular momentum $j$ and
 a label $\alpha$  that distinguishes the different equivalent representations for a given $j$
\begin{equation}
\rho^{\otimes n}=\bigoplus_{j,\alpha} \rho^{j \alpha}.
\end{equation}
The angular momentum takes values \mbox{$j=n/2, n/2-1, \ldots, 1/2(0)$}  for odd (even) $n$ and the  number of equivalent representations for each $j$
is~\cite{deconstruction-2}
\begin{equation}\label{repeated}
\nu_j^n={n \choose n/2-j}\frac{2j+1}{n/2+j+1}\, ,
\end{equation}
That is $ \alpha=1,\ldots,\nu_j^n $. For each block we have~\cite{deconstruction-2}
\begin{eqnarray}\label{deconst}
\tr \rho^{j \alpha}&=&\left(\frac{1-r^2}{4}\right)^{n/2-j} \sum_{k=-j}^j \left(\frac{1-r}{2}\right)^{j-k} \left(\frac{1+r}{2}\right)^{j+k}
\nonumber \\
&\equiv& (2j+1) C^n_j,
\end{eqnarray}
which, of course, is the same for all equivalent irreducible representations,  i.e., independent on the label $\alpha$.
We sketch here the origin of the
factors appearing in \eqref{deconst} (full details can be found in~\cite{deconstruction-2}).
The first factor comes from the contribution from the $n/2-j$ singlets
present in a representation $j$ made up of $n$ spin-1/2 states. The summation term is  the trace of the projection of the
remaining states in the  symmetric subspace with total angular momentum $j$, where we can use the rotational invariance
of the trace to write each state  in diagonal form $\left(\begin{array}{cc}\frac{1+r}{2} & 0 \\0 & \frac{1-r}{2} \end{array}\right)$. This
term simply reads
\begin{eqnarray}
t_j&=&\sum_{k=-j}^j \left(\frac{1-r}{2}\right)^{j-k} \left(\frac{1+r}{2}\right)^{j+k}\nonumber \\
&=&\frac{1}{r}\left[ \left( \frac{1+r}{2} \right)^{2j+1} - \left( \frac{1-r}{2} \right)^{2j+1} \right]
\end{eqnarray}
and hence
\begin{equation}\label{deconstt}
C_j^n=\frac{1}{2j+1}\left(\frac{1-r^2}{4}\right)^{n/2-j} t_j \; .
\end{equation}

Very much in the same way as it happened  in previous sections, the only difference between the diagonal basis of  $ \sigma_1$ and $\sigma_2 $
is the ordering of the  angular momenta couplings. In $ \sigma_1 $ we first couple subspaces $A$ and $B$ and obtain
\begin{equation}\label{rhoAB}
\rho_{AB} = \int dn_1  \rho^{\otimes n}_{1\,  A} \otimes \rho_{1\,  B}^{\otimes m}
= \sum_{\xi_{AB}} C_{j_{AB}}^{n+m} \openone_{\xi_{AB}}\; ,
\end{equation}
where
\begin{equation}
\openone_{\xi_{AB}}=\sum_{M_{AB}}\ketbrad{\xi_{AB} M_{AB}}
\end{equation}
is the projector onto the subspace with quantum numbers $\xi_{AB} = \{j_A,\alpha_A,j_B,\alpha_B,j_{AB}\}$ and $C_{j_{AB}}^{n+m} $ is defined in
Eq.~\eqref{deconst}.
Notice that $C_{j_{AB}}^{n+m}$  depends only on the purity of the state and on the total angular momentum $j_{AB}$.  Notice also that the tensor product of a mixed
state has projections in all subspaces and the blocks are not uniquely determined by the value of  $j_{AB}$, i.e.,  one has to keep track of the labels $j_A$ and $j_B$ as
well. Of course, subspaces with different quantum numbers
$ \xi_{AB} $ are orthogonal, i.e.,  $\tr [\openone_{\xi}\openone_{\xi'}]=\delta_{\xi \xi'}\tr \openone_{\xi} $. When coupling the third
system one plainly adds  the quantum numbers
$\xi_C=\{j_C, \alpha_C\}$.

In the notation we have developed so far, the diagonal bases of $\sigma_1$ and $\sigma_2$  are written as $\mathcal{B}_1=\{\ket{\xi_{AB}\xi_C; JM}\}$ and
$\mathcal{B}_2=\{\ket{\xi_A\xi_{BC}; JM}\}$, respectively. Obviously, each set contains  $2^{2n+m}$  orthonormal states and Eq.~\eqref{s1-s2-int} reads
\begin{eqnarray}\label{s1-s2-int-2}
\sigma_1 &=&  \sum_{\xi_{AB}\xi_C}\sum_{JM} C_{j_{AB}}^{n+m} C_{j_C}^n [\xi_{AB} \xi_C; J  M]  \nonumber \\
\sigma_2 &=&  \sum_{\xi_A\xi_{BC}}\sum_{JM} C_{j_A}^n C_{j_{BC}}^{n+m} [\xi_A  \xi_{BC} ; J  M] .
\end{eqnarray}
We just have to compute the minimum error from the Helstrom formula \eqref{Helstrom-m} for these two states. It is convenient to define the trace-norm term
\begin{equation}\label{tracenorm-mixed}
T=\|\sigma_1-\sigma_2\|,
\end{equation}
so that
\begin{equation}
\PME=\frac{1}{2}\left( 1-\frac{1}{2} T\right).
\end{equation}
To compute $T$ we need to know the unitary matrix $\Lambda$  that transforms $\mathcal{B}_2$ into $\mathcal{B}_1$ or vice versa.
The elements of this unitary are given by the overlaps between the elements of both basis
$\braket{\xi_{AB}\xi_{C}; JM}{\xi'_{A}\xi'_{BC};J'M'}$. We observe that these overlaps are non-vanishing only if  $j_X=j'_X $ ,  $\alpha_X=\alpha'_X $ ($X=A,B,C$) and
$J=J', M=M'$. Furthermore, as mentioned above,  their value does not depend on $M$ or $\alpha_X$, thus, sums over these quantum numbers simply amount to
introduce the corresponding multiplicative factors. Therefore, it is useful to introduce a label containing the quantum numbers  that determine the orthogonal
blocks in $\mathcal{B}_1$ and $\mathcal{B}_2$  that may have non vanishing overlaps, $\xi=\{j_A, j_B,j_C, J\}$ and the corresponding multiplicative factor
\begin{equation}\label{multiplicative-factor}
\gamma_{\xi}=\nu_{j_A}^{n}\nu_{j_B}^{m}\nu_{j_C}^{n}(2J+1),
\end{equation}
where  $\nu^n_j$ is given in Eq.~\eqref{repeated}.
Eq.~\eqref{tracenorm-mixed} then reads
\begin{eqnarray}\label{T}
T&=&\sum_\xi \gamma_\xi T^{\xi}= \sum_\xi \gamma_\xi \|\sigma^{(\xi)}_1-
\Lambda^{(\xi)} \sigma^{(\xi)}_2 {\Lambda^{(\xi)}}^{T} \|,
\end{eqnarray}
where the explicit expressions of the matrix elements are
\begin{eqnarray}\label{sigma-xi}
 (\sigma_1^{(\xi)})_{j_{AB}j'_{AB}}&=&\delta_{j_{AB}j'_{AB}} C_{j_{AB}}^{n+m} C_{j_C}^n \nonumber \\
  (\sigma_2^{(\xi)})_{j_{BC}j'_{BC}}&=&\delta_{j_{BC}j'_{BC}} C_{j_{A}}^{n} C_{j_{BC}}^{n+m}
\end{eqnarray}
and
\begin{equation}\label{lambda}
\Lambda^{(\xi)}_{j_{AB},j_{BC}}=\braket{\xi, j_{AB}}{\xi, j_{BC}}.
\end{equation}
Recall that the overlap \eqref{lambda} is independent of the quantum number labelling the equivalent representations (recall also that it is independent of $M$) and therefore is
given by Eq.~\eqref{6j}.

The computation of the minimum error probability  reduces  to a sum of trace-norms of  small size Helstrom matrices that have dimensions
of the allowed values of  $j_{AB}$ and $j_{BC}$ for given $\xi=\{j_A,j_B, j_C, J\}$. Hence
\begin{equation}\label{pme-general}
\PME = \frac{1}{2} \left( 1 - \frac{1}{2}\sum_{\xi} \gamma_\xi T^{\xi}\right)
\end{equation}
and this computation can be done very efficiently.

We would like to show the analytical results for  the simplest case of having just one state at each port, i.e. when $n=m=1$. In this situation we have fixed values
$ j_A=j_B=j_C=1/2 $, the total angular momentum can be $ J=3/2,1/2 $ and $ j_{AB}=1,0 $ (and similarly for $ j_{BC} $). Here there is no degeneracy,
the number of equivalent representations defined in Eq.~\eqref{repeated}  is one,
and therefore the multiplicative factor \eqref{multiplicative-factor}  simply reads $\gamma_\xi=2J+1$.
The only relevant quantum number in this case is $\xi=J$,  as all the other are fixed, and we do not need to write them explicitly.  The minimum error probability is then
\begin{equation}\label{mixedn1m1}
P^{\rm{ME}}=\frac{1}{2}\Big(1-\frac{1}{2}\sum_{J=1/2}^{3/2}(2J+1)\|\sigma_1^{(J)}-\Lambda^{(J)}\sigma_2^{(J)}{\Lambda^{(J)}}^T\|\Big)
\end{equation}
The term of the sum corresponding to $ J=3/2 $ vanishes since it corresponds to the projection of $ \sigma_{1,2} $ onto the completely symmetric
subspace, which is identical for both states.
Indeed, in  this subspace $\sigma_1^{(3/2)}=\sigma_2^{(3/2)}=C^2_1 C^1_{1/2}=(3+r^2)/24$, where we have used Eq.~\eqref{deconstt},
and from Eq.~\eqref{lambda} we obtain $\Lambda^{(3/2)}=1$. In the subspace $J=1/2$ we have
\begin{eqnarray}
		 \sigma_1^{(1/2)} =  \sigma_2^{(1/2)}  &=&
		\begin{pmatrix}
			C^2_1 C^1_{1/2} & 0 \\
			0 & C^2_0 C^1_{1/2}
		\end{pmatrix} \nonumber \\
		&= &
		\begin{pmatrix}
			\frac{1}{24}\left(3+r^2\right) & 0 \\
			0 & \frac{1}{8}\left(1-r^2\right)
		\end{pmatrix}
\end{eqnarray}
and
\begin{equation}
	\Lambda^{(1/2)} =
		\begin{pmatrix}
			\frac{1}{2} & \frac{\sqrt{3}}{2} \\
			\frac{\sqrt{3}}{2} & -\frac{1}{2}
		\end{pmatrix} \, .
		\label{ss3}
\end{equation}
 Plugging these expressions  into \eqref{mixedn1m1} we obtain the minimum error probability of one copy state
\begin{equation}\label{1x1x1}
P^{\rm{ME}}=\frac{1}{2}\left(1-\frac{r^2}{2\sqrt{3}}\right).
\end{equation}
As expected, when $r\to 1$ we recover the pure state value ~\eqref{minerr-tot-2}.
\begin{figure}[ht]
\setlength{\unitlength}{.9cm}
\begin{picture}(9,4.7)
\put(-.3,0){
\includegraphics[width=7.7cm]{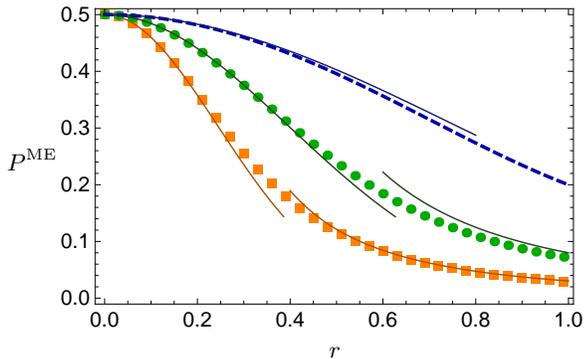}
}
\end{picture}
\caption[]{\label{fig:fig1} (color online) Error probability $ P^{\rm{ME}} $ for $ n=m=3 $ (blue dashed line)$,11$ (green circles) and $ 29 $ (yellow squares) versus purity. The fit $ \PME \simeq 0.882/(n r^2) $ in the regime of high purities for $ n=11 $ and $ n=29 $ and the Gaussian approximation $\PME\simeq 1/2 \exp[-n r^2/(2\sqrt{3})]$ in the regime of low purities for all cases is represented (solid lines).}
\end{figure}

\begin{figure}[ht]
\setlength{\unitlength}{.9cm}
\begin{picture}(9,4.7)
\put(-.3,0){\includegraphics[width=7.7cm]{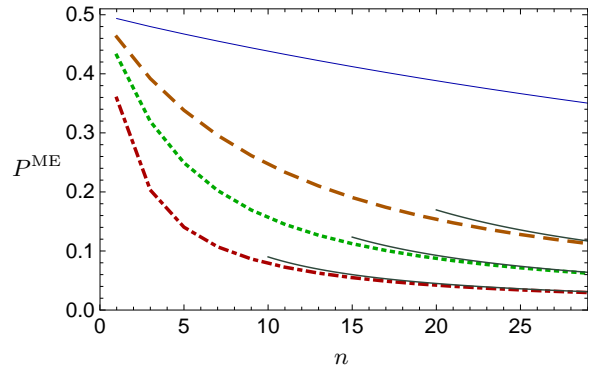}}
\end{picture}
\caption[]{\label{fig:fig2} (color online) Error probability $ P^{\rm{ME}} $ for $ r=0.2 $ (blue thin solid line), $ r=0.5 $ (brown dashed line), $ r=0.7 $ (green dotted line)
 and $ r=1 $ (red dot-dashed line) versus $ n $ ($ n=m $ is assumed). Numerical points have been joined for an easier visualization.  The approximation $0.882/(nr^2)$  is also represented (thin solid lines).}
\end{figure}

Numerical results  of the minimum error probability as a function of the purity of the input states for the symmetric case
 $n=m$ are depicted in  Fig.~\ref{fig:fig1}. One sees that for low values of $n$ ($n\lesssim 3$) the dependence on the purity is not very marked, the curves
are concave almost in the whole range of the purity. For larger $n$ however there is an interval of purities where the behavior changes quite
significantly.  For, e.g., $n=29$ the inflection point occurs at $r\approx 0.3$.  At very large values of $n$ one expects a step-like shape with an inflection point
approaching $r=0$ because the probability of error remains very small for $r\neq0$ and is strictly 1/2 at $r=0$.  The shape of the curves is explained by the existence of two distinct  regimes. For high purities the probability of error is well fitted by a linear function in the inverse of the number of copies. We get  $\PME\simeq 0.88/(nr^2)$ where the value $0.88$ coincides with the analytical value computed for pure states
Eq.~\eqref{pure-assymptotic}.  Of course,  this  approximation cannot be valid for low purities.  In this range of low purity the minimum error probability is very well approximated by the Gaussian function $\PME\simeq 1/2 \exp[-n r^2/(2\sqrt{3})]$, where we have taken the argument of the exponential  from the exponentiation of the exact $1\times1 \times 1$ case
\eqref{1x1x1}.  This approximation works for purities in the interval of  the width of the Gaussian, i.e., up to  $\sim1/\sqrt{n}$.  Therefore, as $n$ increases  the asymptotic approximation  $\PME\propto 1/(nr^2)$ extends its validity to almost the whole range of purities, and the expected jump discontinuity develops in $r=0$ as $n\to \infty$.
 Similar information is depicted in Fig.~\ref{fig:fig2}, where the error probability is plotted as function of the number of copies $n$ for different purities. We have superimposed the asymptotic result, which  is seen to yield a very good approximation to the exact error probability already for $n\gtrsim 20$.

\section{Asymptotic $n\times1\times n$}\label{sec:mixed-limits}

As in previous sections, it is interesting to study the performance of the machine in the asymptotic regimes. A particularly important instance where it is possible
to obtain closed expressions is the case when the number of copies at the program ports is asymptotically large and there is one state at the data port.
We show how to  compute the leading order and  sketch the generalizations needed to obtain the subleading term.

Observe first that
$j_{AB}$  can only take the values $j_{AB}=j_A\pm1/2$ and similarly for $j_{BC}$. Therefore  $\sigma_{1,2}^{(\xi)}$ are $2\times2$ matrices (except in the
extremal case of $J=j_A+j_C+1/2$ which is one-dimensional).  It is useful to write
\begin{equation}\label{sigma-j}
\sigma(j)=C^{n}_{j_{A}}C^{n}_{j_{C}} \begin{pmatrix}  R_{+}(j) & 0 \\
                                                                                              0       & R_{-}(j),
                                                 		\end{pmatrix}
\end{equation}
with
\begin{equation}
R_{\pm}(j)=\frac{C_{j\pm1/2}^{n+1}}{C_j^n}.
\end{equation}
With this definition one simply has [see Eq.\eqref{sigma-xi}]:
\begin{equation}
\sigma_1^{(\xi)}=\sigma(j_A) \ \ \ \ \ \mathrm{and} \ \ \ \ \ \sigma_2^{(\xi)}=\sigma(j_C).
\end{equation}
We further notice that for large $n$
\begin{eqnarray}\label{gauss-approx}
\nu_j^n C_j^n & \approx& \frac{1}{n/2+j+1} \frac{1+r}{2r}\nonumber\\
& &\ \ \times \sqrt{\frac{2}{n \pi (1-r^2)}}
\exp \left[-n\frac{(2j/n-r)^2}{2(1-r^2)}\right].
\end{eqnarray}
Defining $y=2j/n$  and using the  Euler-Maclaurin summation
formula,
we have  for a generic function $f(j)$
\begin{equation}\label{gauss-approx-2}
\kern -0.2 em
\sum_j \nu_j^n C_j^n f(j)   \approx \frac{1+r}{2r} \int_{-\infty}^{\infty}\frac{dy  \, G_n(y) }{n/2+n y/2+1}
f(\frac{ny}{2}) ,
\end{equation}
\begin{figure}[ht]
\setlength{\unitlength}{.9cm}
\begin{picture}(9,4.7)
\put(-.3,0){\includegraphics[width=7.7cm]{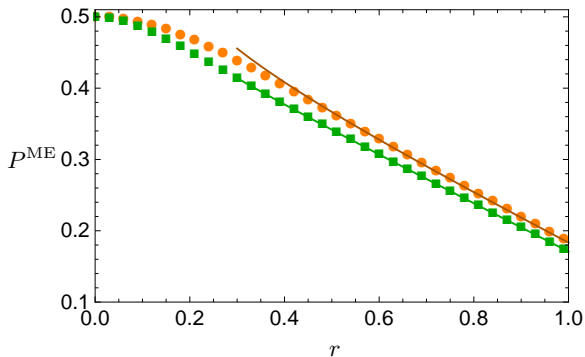}}
\end{picture}
\caption[]{\label{fig:fig3} (color online) Error probability $ P^{\rm{ME}} $ for $ n=20 $ (yellow circles) and $ n=79 $ (green squares) versus purity. The asymptotic behaviour given by Eq.~\eqref{mixed-subleading} is represented for both cases.}
\end{figure}
where we have extended  limits of integration from (0,1)  to $(-\infty,\infty)$, which is legitimate for large $n$, and defined
\begin{equation}
G_n (y)=\sqrt{\frac{n}{2 \pi (1-r^2)}}\exp \left[-n\frac{(y-r)^2}{2(1-r^2)}\right],
\end{equation}
a Gaussian distribution centered at $y=r$  with variance $\sigma^2=(1-r^2)/n$.
Notice that at leading order, $n\to \infty$,  $G_{\infty}\approx \delta(y-r)$ and hence
\begin{equation}
\sum_j \nu_j^n C_j^n f(j)   \approx \frac{1}{n r}  f(\frac{nr}{2}).
\end{equation}
Notice also that at this order

\begin{equation}\label{rpm-leading}
R_{\pm}(j)\approx R_{\pm}(\frac{nr}{2})=\frac{1\pm r}{2}.
\end{equation}
There only remains  to compute the unitary matrix Eq.~\eqref{lambda}. Observe that the  total angular momentum takes values
$J=|j_A -j_C|+1/2+k$ with $k=0,1,\ldots,2 \min\{j_A,j_C\}$.  The leading order is rather easy to write (the subleading term,
although  straightforward, is far more involved,  and we will not show it here).  At this order we have
$J=1/2+k$ and  $k=0,1,\ldots, n r$ and the matrix elements computed from \eqref{6j}  yield
\begin{equation}\label{lambda-leading}
\Lambda^{(\xi)}=\frac{1}{nr}\begin{pmatrix} k & \sqrt{(nr)^2-k^2} \\
                                                                        \sqrt{(nr)^2-k^2} & -k
                                              \end{pmatrix}\, .
\end{equation}
Plugging  Eqs.(\ref{sigma-j}-\ref{lambda-leading}) into Eq.~\eqref{T} one gets
\begin{equation}
T\simeq \sum_{k=0}^{nr} 2k\times \frac{2}{n^3 r^2} \sqrt{(nr)^2-k^2}
\end{equation}
where the sum over $j_A$ and $j_C$ has been trivially performed by substituting  their central value $n r/2$ in the summand and the only remaining multiplicative
of $\gamma_\xi$ [cf. Eq.~\eqref{multiplicative-factor}]  is $2J+1\simeq 2k$.  Finally, defining $x\equiv k/nr$ and using the Euler Mac-Laurin approximation as
in Eq.~\eqref{SME-sum} we obtain
\begin{equation}
T\simeq 4 r\int_0^1 dx \, x\sqrt{1-x^2}=\frac{4r}{3}\, ,
\end{equation}
and hence
\begin{equation}\label{mixed-leading}
\PME\simeq \frac{1}{2}-\frac{r}{3}\, ,
\end{equation}
which obviously coincides with the pure state result Eq.~\eqref{SME-sum} for $m=1$ and $r\to 1$.

As for the computation of the next-to-leading order,
 the integrals approximating the sums over $j_A$ and $j_C$  have to incorporate the fluctuations around the central value, i.e.,
  one defines  $j_A=\frac{n}{2}(r+\eta_A)$ and $j_C=\frac{n}{2}(r+\eta_C)$, where the variables $\eta_X$ have effective dimension $n^{-1/2}$.
Then one can expand
the matrix elements of $\sigma_{1,2}$,
$\Lambda$, and the terms of $\nu_j^{n}$ present in Eq.~\eqref{gauss-approx-2} taking into account the
effective dimensionality of all the terms [notice that $k\to n(r+\eta) x$, where the integration range of  $x$ is $(0,1)$] .
 One then performs the sum in $k$ by means of the Euler-Maclaurin
summation formula as before. Finally one computes the  integration in $j_{A/B}$ taking into account that
range of the  variables $\eta_{A/B}$  can be taken to be  $(-\infty,\infty)$.
After a somewhat lengthy calculation we obtain
\begin{equation}\label{mixed-subleading}
\PME\simeq \frac{1}{2}-\frac{r}{3}+\frac{1}{3n r}\, .
\end{equation}
Notice that the limit $r=0$ is singular and not surprisingly the  expansion breaks down for purities of order $1/n$. As it should, the error probability \eqref{mixed-subleading}
increases monotonically with the purity.

In Fig.~\ref{fig:fig3} we plot the error probability as a function of the purity for $n=20$ and $n=79$. One sees that the asymptotic expression
 \eqref{mixed-subleading} approximates very well the minimum error probability even for small number of copies . For larger $n$, e.g.,  for $n=79$
 the approximation works extremely well down to values below $r=0.3$.

We finish this section by showing that the leading term \eqref{mixed-leading}  coincides with the average error of
a device that first estimates the mixed states at the program ports and afterwards does the usual minimum error discrimination of the
data state. From the Helstrom formula~\eqref{Helstrom-m}
particularized for mixed qubit states one has
\begin{equation}
\PME=\Bigr\langle \frac{1}{2}\left( 1-\frac{1}{2}|\vec{r}_1-\vec{r}_2| \right) \Bigr\rangle
\end{equation}
where the average is taken over all possible orientations of the Bloch vectors $\vec{r_1}$ and $\vec{r}_2$. For equal purity states it simply reads
\begin{equation}
\PME =\frac{1}{2}\left(1-\frac{r}{2}\int_0^\pi d\theta  \sin\theta \sin\theta/2\right) =\frac{1}{2}-\frac{r}{3}.
\end{equation}

\section{Fully Universal discrimination machine}

Let us finally address the  fully universal discrimination machine, i.e., a machine that distinguishes states from which nothing is assumed to be known, not
even its purity.  For this type of machine, we  need to specify
\begin{figure}[ht]
\setlength{\unitlength}{.9cm}
\begin{picture}(9,4.7)
\put(-.3,0){\includegraphics[width=7.7cm]{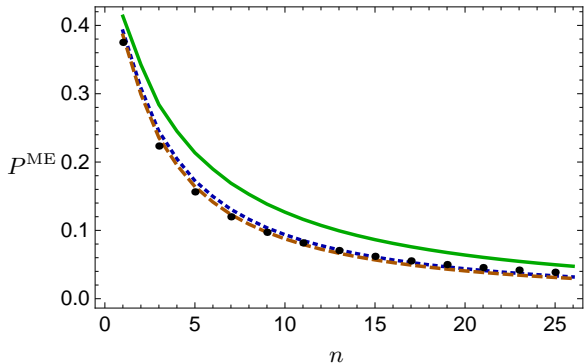}}
\end{picture}
\caption[]{\label{fig:fig4} (color online) Error probability $ P^{\rm{ME}} $ for hard-sphere (green solid line), Bures (blue dotted line) and Chernoff (red dashed line) priors versus $ n $ ($ n=m $ is assumed). The points correspond to the error probability for a fixed $ r=0.9 $; its proximity to the Chernoff curve exposes the fact that this prior gives larger weights to states of high purity.}
\end{figure}
a prior distribution for the purity.
While the isotropy of the angular variables yields a unique uniform
distribution for the angular variables, the Haar measure on the 2-sphere used in previous sections, the corresponding expression for a fully unbiased distribution of
the purity $w(r)$ is not uniquely determined.   This is a longstanding issue, and several priors haven been suggested depending on the assumptions
made~\cite{petz,geometry}. Here we will  not stick to a particular distribution, rather we will show results for three reasonable distributions. The actual values of the the  probability of error may depend on the chosen prior, but the overall performance is seen to be very similar.

The most straightforward, but perhaps not very well grounded, choice is that of the distribution of a hard-sphere $w(r)\propto r^2$, i.e., a normalized integration measure given by
\begin{equation}\label{hard}
d\rho^\mathrm{HS} = 3 r^2 dr \frac{d\Omega}{4\pi}\, .
\end{equation}

The Bures distribution is far better motivated. It corresponds to the volume element induced by the fidelity distance~\cite{zyc}. It is monotonically decreasing under coarse graining~\cite{petz} and it has been argued that it corresponds to maximal randomness of the signal states~\cite{hall}.  In this case one has $w(r)\propto r^2/\sqrt{1-r^2}$. Notice that this distribution assigns larger weights to pure states, as their distinguishability  in terms of the fidelity is larger than that of mixed states.
The integration measure reads
\begin{equation}\label{bures}
d\rho^{\mathrm{Bu} }= \frac{4}{\pi}\frac{r^2}{\sqrt{1-r^2}} dr \frac{d\Omega}{4\pi}\, .
\end{equation}

Lastly, we also consider the recently proposed Chernoff distribution~\cite{chernoff}. It is the prior induced by the Chernoff distance, which has a clear
operational meaning in terms of the distinguishability between states. By construction it is  monotonically decreasing under coarse graining. This measure
assigns  even larger weights to states of high purity and lower to the very mixed ones. This assignment is again based on the  distinguishability properties, but in terms of the asymptotic behavior of the error probability.  The measure can be written as~\cite{chernoff}
\begin{equation}\label{chernoff}
d\rho^{\mathrm{Ch} }= \frac{1}{\pi-2}\frac{\left(\sqrt{1+r}-\sqrt{1-r}\right)^2}{\sqrt{1-r^2}} dr \frac{d\Omega}{4\pi}\, .
\end{equation}

The effective states we have to discriminate are
\begin{eqnarray}\label{s-universal}
\Sigma_k &= & \int d\rho_1 d\rho_2 \rho^{\otimes n}_{1\,  A} \otimes \rho_{k\,  B}^{\otimes m} \otimes \rho^{\otimes n}_{2\,  C}  \ (k=1,2),
\end{eqnarray}
where $d\rho_k$ takes the expressions of the measures (\ref{hard}-\ref{chernoff}). Note that the block structure of the states is the same as before, as it only depends on the permutation invariance  of the input states,  which remains untouched. Further, we can use rotational invariance in the same fashion as in Eqs.~\eqref{rhoAB} and~\eqref{s1-s2-int-2}.  Therefore,  here it is only required  to compute the average of the coefficients $C_j^{n}$ in Eq.~\eqref{deconst} according to priors (\ref{hard}-\ref{chernoff}). To calculate
the minimum error probability of this fully universal machine one simply uses Eq.~\eqref{pme-general}  for the states~\eqref{s1-s2-int-2} with the averaged coefficients $\aver{C_j^n}$ computed in
Appendix~\ref{ap:averages}.

In Fig.~\ref{fig:fig4} we present the minimum error probability of the fully universal machine for the three priors discussed for an equal number of program and data states up to $n=m=26$.
As anticipated, the smaller average error corresponds to the  Chernoff distance, because states with higher purity are assigned a larger weight, and  these are easier to discriminate.  The probability of error, as somehow expected, is inversely proportional to the number of copies, and attains very similar values than for the discrimination of states with fixed known purity of the order of $r\sim 0.9$.

\section{Conclusions}

We have studied the problem of programmable discrimination of two unknown general qubit states when multiple copies of the states are provided.
For pure states we have obtained the  optimal unambiguous discrimination  and minimum-error probabilities, Eqs.~\eqref{ua-nm} and  \eqref{min-nm}, respectively. Some results along these lines can be found in~\cite{he-bergou},  however no closed expressions were given there. Knowing the error in the asymptotic regimes is a very relevant information as it allows to assess and compare the performance of devices in a way which is independent
on the number of copies. We have obtained analytical expressions for the leading and sub-leading terms in several cases of interest. As could be anticipated,  when the number of copies at the program ports is asymptotically large at leading order we recover the average of the usual discrimination problem of known states~Eqs. \eqref{SUA-known-average} and \eqref{PME-int}. When the data port is loaded with an asymptotically large number of copies, we recover the state comparison averaged error Eqs.~\eqref{eq:UA-averaged} and \eqref{eq:ME-averaged}. These cases correspond to measure and discriminate protocols, where the measurement unveils the classical information about the states.

 We have addressed for the first time the programmable discrimination of copies of mixed states.
We have obtained the minimum-error probability when the ports are loaded with
copies of qubits of known purity, Eq.~\eqref{pme-general}.
We have assumed that all states have the same purity. This would, e.g., correspond to a scenario where all the initially pure data and program states are subject to the same depolarizing noise before entering the machine.
 Closed analytical results for a small number of copies can be obtained and efficiently computable expressions for a fairly large number of copies are given. The asymptotic analytical results show very good agreement with the numerics.
The latter show a characteristic $1/N$ dependence with the number $N$ of available copies ---in contrast to the  usual exponential decay found in standard (non-universal) state discrimination--- and provide a very good approximation already for relatively low number of copies when the states have high purity.
For very mixed states the error probability has a drastically different behavior.  Logically, in both cases the error probability monotonically decreases with increasing purity $r$, but in the low purity regime the dependence is much less
pronounced.  The range of purities exhibiting this behavior shrinks as the number of copies increases, and the characteristic $1/N$ behavior of the asymptotic regime extends its validity
over almost  the whole range of purities.



Finally we have studied the fully universal discrimination machine, a device that takes in states of which nothing is known, i.e.,  not even its purity. We compute the minimum error probability for three reasonable prior distributions of the purity: hard-sphere, Bures and Chernoff (see Fig.~\ref{fig:fig4}). The latter  is seen to give the lowest error probability.
 This comes as no surprise, since the Chernoff distribution assigns larger weights to pure states (because they are better distinguished). Our results also indicate that the fully universal discrimination states yields an error probability comparable to the discrimination of states of known purity,  being that remarkably large ($r\sim 0.9$).


\acknowledgments
We thank J. Bergou for their contributions in the earlier stages of this work.
We acknowledge financial support from:
the Spanish MICINN, through the Ram\'on y Cajal program~(JC), FPI grant BES-2009-028117 (GS), contract FIS2008-01236, and project QOIT
(CONSOLIDER2006-00019); from the Generalitat de
Catalunya CIRIT, contract  2009SGR-0985.
We thank M. Hayashi for pointing out a discrepancy with the wrong~Eq. (A2) in the published version of our paper.

\appendix
\section{Arbitrary copies of pure states}\label{na-nb-nc}
In this appendix we present the unambiguous discrimination and minimum error probabilities when the number of copies $ n_A,n_B,n_C $  loaded at the machine ports is completely  arbitrary. Note that in this case the global states $ \sigma_1 $ and $ \sigma_2 $ [cf. Eq.~\eqref{rho1-rho2}] may
have different dimensions, for  $ d_1 = (\na+\nb+1)(\nc+1) $ is in general not equal to  $ d_2 = (\na+1)(\nb+\nc+1) $.  One can easily convince oneself that the support  of the state with smallest dimension is always contained in the support of the other, and hence the problem can be solved in very much the same way as in the main text simply taking into account that the error probabilities now only contain contributions from the intersection of the supports. Without loss of generality we
can assume from now on that $\na\geq \nc$.
As discussed in the main text, the error probabilities are computed by adding the pairwise contributions of the state bases in the common support, the main difference being that $ \sigma_1 $ and $ \sigma_2 $ do not have equal coefficients in front of the projectors and hence the prior probabilities of each pair of states are different. Also, the overlaps in Eq.~\eqref{6j}  will have a slightly more complicated expression. Here we have $ j_A=\na/2 $, $ j_B=\nb/2 $, $ j_C=\nc/2 $, $ j_{AB}=(\na+\nb)/2 $ and $ j_{BC}=(\nb+\nc)/2$. The minimum $ J $ available for $ \sigma_1 $ is $ j_B+j_A-j_C \equiv J_{min}^1$, and   $ |j_B+j_C-j_A| \equiv J_{min}^2 $ for $ \sigma_2 $. The maximum angular momentum $ j_A+j_B+j_C \equiv J_{max} $ is reachable for both states. 
For equal prior probabilities for $ \sigma_1 $ and $ \sigma_2 $, we can write
\begin{eqnarray}
\frac{1}{2} \sigma_1 &=& \sum_{J=J_{min}^1}^{J_{max}}\sum_{M=-J}^J p_J\, \pi^{1}_J { [j_{AB} ; J M] } , \\
\frac{1}{2} \sigma_2 &=& \sum_{J=J_{min}^2}^{J_{max}}\sum_{M=-J}^J p_J\, \pi^{2}_J { [j_{BC} ; J M] } ,
\end{eqnarray}
where  
$
p_J={1\over 2}\left({1\over d_1}+{1\over d_2}\right)$, $\pi^{1}_J={1\over 2p_J\,d_1}$, $\pi^{2}_J={1\over 2p_J\,d_2}
$
for $J_{min}^1\le J\le J_{max}$, whereas
$
p_J={1\over 2d_2}$, $\pi^{1}_J=0$, $\pi^{2}_J=1
$
for $J_{min}^2\le J<J_{min}^1$. We can view $ p_J $ as the probability of projecting on subspace labeled by a given $ J $ and $ M $, and $\pi^{1}_J$, $\pi^{2}_J$ as the prior probabilities that, on that subspace, the unknown state be $[j_{AB} ; J M]$ or $[j_{BC} ; J M]$.
If the condition
\begin{equation}\label{uaineq}
{c^2_{J} \over 1+c^2_{J}}\le \pi_J^{AB}\le{1 \over 1+c^2_{J}} \, ,
\end{equation}
where $c_J=|\langle j_{AB} ; J M | j_{BC} ; J M \rangle|$ is given by Eq.~\eqref{6j}, holds, then the probability of obtaining an inconclusive answer when we finally discriminate between $[j_{AB} ; J M]$ and $[j_{BC} ; J M]$ is~\cite{uaasym} $P^{?}_J=2\sqrt{\pi_J^1 \pi_J^2} c_J $.
If Eq.~\eqref{uaineq} is satisfied for $ {\hat J}=J_{max}-1 $, then it will be satisfied all over this range of $ J $, since $ c_J $ is a monotonically increasing function of $ J $. The overlap $ c_{\hat J} $ has the very simple form
\begin{equation}
c^2_{\hat J}={n_A n_C\over (n_A+n_B)(n_B+n_C)}.
\end{equation}
Thus, Eq.~\eqref{uaineq} is equivalent to
\begin{eqnarray}
{n_A n_C\over (n_A+n_B)(n_B+n_C)}
&\le &
{(n_A+n_B+1)(n_C+1)\over(n_B+n_C+1)(n_A+1)} \nonumber \\
&\le & {(n_A+n_B)(n_B+n_C)\over n_A n_C} \,,
\end{eqnarray}
which is clearly true. Eq.~\eqref{uaineq} does not hold if $ J=J_{max} $, for which we have $P^{?}_{J_{max}}=1$. 
Notice that since no error is made for $J_{min}^2\le J<J_{min}^1$, for $\pi^{1}_J=0$, the total
inconclusive probability  reads $P^{?}=\sum_{J=J_{min}^1}^{J_{max}}
p_J\, (2J+1)P^{?}_J$, which has the explicit expression
\begin{widetext}
\begin{equation}
P^{?} = \frac{1}{2} \left(\frac{1}{\sqrt{d_1}}-\frac{1}{\sqrt{d_2}}\right)^2 d_{ABC}  + \frac{1}{\sqrt{d_1d_2}} \sum_{k=0}^{n_C}
(n_A+n_B-n_C+2k+1)
\sqrt{
{
\scriptsize
\begin{pmatrix} n_A+n_B-n_C+k \\  n_B \end{pmatrix}
\begin{pmatrix} n_B+k \\  n_B \end{pmatrix}
              \over
\begin{pmatrix} n_A+n_B \\ n_B \end{pmatrix}
\begin{pmatrix} n_C+n_B \\ n_B \end{pmatrix}
}
} ,
\end{equation}
\end{widetext}
where $ d_{ABC}=n_A+n_B+n_C+1 $. Note that when $\na = \nc$  the term proportional to $ d_{ABC} $ disappears and the square root term simplifies, so we recover the closed form given in the main text [cf. Eq.~\eqref{ua-nm}].

The minimum error probability can be computed entirely along the same lines. For a pair of states we have~\cite{uaasym} $P_J^{\rm ME}={1\over2}\left(1-\sqrt{1-4\pi^{1}_J\pi^{2}_Jc_J^2} \right)$, and the total error probability reads
\begin{widetext}
\begin{equation}\label{measym}
P^{\rm ME} = {1\over4}\left\{1+{d_1\over d_2}-{d_1+d_2\over d_1d_2}\sum_{k=0}^{n_C}
(n_A+n_B-n_C+2k+1)
\sqrt{
1-4{d_1d_2\over(d_1+d_2)^2}
{
\scriptsize
\begin{pmatrix} n_A+n_B-n_C+k \\  n_B \end{pmatrix}
\begin{pmatrix} n_B+k \\  n_B \end{pmatrix}
              \over
\begin{pmatrix} n_A+n_B \\ n_B \end{pmatrix}
\begin{pmatrix} n_C+n_B \\ n_B \end{pmatrix}
}
}
\right\} .
\end{equation}
\end{widetext}
This expression coincides with Eq. (31) of~\cite{masahito}.

\section{Averaged $C_j^n$ coefficients}\label{ap:averages}

Here we compute the average of the coefficients [see Eq.~\eqref{deconst}]
\begin{eqnarray}\label{cnj}
C^n_j&=&\frac{1}{2j+1}\left(\frac{1-r^2}{4}\right)^{n/2-j} \nonumber \\
&&\times \sum_{k=-j}^j \left(\frac{1-r}{2}\right)^{j-k} \left(\frac{1+r}{2}\right)^{j+k}
\end{eqnarray}
for the hard sphere, Bures and Chernoff priors, Eqs.~\mbox{(\ref{hard}-\ref{chernoff})}, considered in the fully universal discrimination machine.

For the hard sphere prior we have
\begin{eqnarray}
\aver{C_j^n}_\mathrm{HS} &=& 3 \int C_j^n r^2 dr \nonumber \\
&=& 6 \frac{\Gamma(n/2+j+2) \Gamma(n/2-j+1)}{\Gamma(n+4)} \; .
\end{eqnarray}

The Bures distribution yields
\begin{eqnarray}
\aver{C_j^n}_{\mathrm{Bu}} & =& \frac{4}{\pi} \int C_j^n \frac{r^2}{\sqrt{1-r^2}} dr \nonumber  \\
&=& \frac{4}{\pi}\frac{\Gamma(n/2+j+3/2) \Gamma(n/2-j+1/2)}{\Gamma(n+3)} \; .
\end{eqnarray}

The averages for the  Chernoff prior are a bit more involved, but  still can be given in a closed form as
\begin{eqnarray}
\aver{C_j^n}_{\mathrm{Ch}} &=& \frac{1}{\pi-2} \int C_j^n \frac{\left(\sqrt{1+r}-\sqrt{1-r}\right)^2}{\sqrt{1-r^2}} dr \nonumber \\
&=&  \frac{2}{(\pi-2)(2j+1)} \sum_{m=-j}^j \left[ B_{1/2} \left( \tfrac{n+1-2m}{2},\tfrac{n+1+2m}{2}\right) \right.\nonumber \\
&&\phantom{xxxxxxxxxxx}\left. - 2 B_{1/2} \left( \tfrac{n-2m+2}{2},\tfrac{n+2m+2}{2}\right)\right]
\end{eqnarray}
%
where $ B_x(a,b)=\int_0^x t^{a-1} (1-t)^{b-1} dt $ is the incomplete beta function~\cite{abra}.


\end{document}